\documentclass[pre,reprint,superscriptaddress,amsmath,amssymb,aps,showpacs, floatfix,]{revtex4-1}

\usepackage{graphicx}
\usepackage{dcolumn}
\usepackage{bm}
\usepackage{epstopdf}
\usepackage{color}
\usepackage{booktabs}
\usepackage{comment}
\usepackage{psfrag}

\def\beq{\begin{equation}}
\def\eeq{\end{equation}}
\def\bea{\begin{eqnarray}}
\def\eea{\end{eqnarray}}

\def\openone{\leavevmode\hbox{\small1\kern-3.3pt\normalsize1}}

\begin{document}

\title{Probing models of information spreading in social networks}

\author{J. Zoller}
\affiliation{Institut f\"ur Quanteninformationsverarbeitung \& IQST, 
Universit\"at Ulm, 89069 Ulm, Germany}

\author{S. Montangero}
\affiliation{Institut f\"ur Quanteninformationsverarbeitung \& IQST, 
Universit\"at Ulm, 89069 Ulm, Germany}

\date{\today}
\pacs{64.60.aq,89.65.-s,95.75.Wx}

\begin{abstract}
We apply signal processing analysis to the information spreading in scale-free network. To reproduce typical behaviors obtained from the analysis of information spreading in the world wide web we use a modified SIS model where synergy effects and influential nodes are taken into account. 
This model depends on a single free parameter that characterize the memory-time of the spreading process. We show that by means of fractal analysis it is possible -from aggregated easily accessible data- to gain information on  the memory time of the underlying mechanism driving the information spreading process. 
\end{abstract}

\maketitle

The study of networks is now fast developing 
due also to its possible impact in different fields with practical applications as communications, healthcare politics, marketing and social sciences~\cite{BarabasiNP11}. Indeed, in the last decades the impressive development of the field has 
unveiled many universal properties of scale-free networks and shown that typical real-world networks 
display the expect theoretical properties~\cite{BarabasiRMP02,castellano09}.
Understanding information flows in social networks is one of the open problems in network science: it is highly non trivial as it might depend on network geometry, propagation rules and system parameters. Different studies addressed this problem either by 
means of observation of real data~\cite{leskovec07,lewisPNAS12,Liben-nowell2008,viswanath09}, 
introducing metrics to quantify network social dynamics and influential spreaders~\cite{pala12,Lind07,Gallos12}, 
validating different hypothesis by means of artificially  structured networks~\cite{centolaSc10}, or putting forward different theoretical 
models of spreading processes~\cite{moreno04,youngPNAS11,rechePRL11, shaoPRL09,Liu2012}.
{Moreover social dynamics have been explored by statistical methods~\cite{Araujo10,Moreira06}.}
The efforts made in the aforementioned analysis is enormous as usually a huge amount of data have to be processed and 
extensive numerical simulations on large network sizes have to be performed to support and verify theoretical analysis.  
Indeed, this is a typical problem encountered in complex system analysis, where rarely microscopic models of the actions of the 
elementary constituents or agents are known in details (being brain cells, ants, or market agents); 
nor simulations of the whole system evolution based on first principles are possible: 
For example, simulating the brain activity including a detailed description of each single neuron, 
has been considered -since very recently- practically impossible~\cite{hbp}.
A possible approach to overcome this limitations adopted in standard complex system analysis is to 
study time correlations of signals extracted from a complex system or part of it --being brain, social or stock market activities--  
and inferring important information on the global status of the system or on the ongoing processes~\cite{grigo10}.
Along these lines, one of us introduced a tool to investigate the properties of time series
extracted from the evolution of social networks and applied it to 
the analysis of the world wide web~\cite{montangero12}.  In particular, the method introduced in Ref.~\onlinecite{montangero12} is based on 
the working hypothesis that the correlations present in time data series are representative of the activities of the underlying communities, 
and thus by studying them it is possible to indirectly infer properties of the agents themselves and of the interactions between them. 
It has been shown that correlations can be quantified by means of the 
fractal analysis of the signal~\cite{sachrajdaPRL98}, and argued that 
a fractal signal corresponds to a strong active community, possibly very influential 
and with high probability of lasting for a long period of time. 

In Ref.~\cite{montangero12} this approach has been introduced heuristically and applied 
to the analysis of real-world time series, i.e. occurrences of keywords in the world wide web; 
in this work we follow a bottom-up approach to test the working hypothesis and to 
look for theoretical models of the fundamental mechanism responsible 
for the creation of such correlations compatible with the experimental data. 
\begin{figure}[t]
\includegraphics[width=0.25\textwidth]{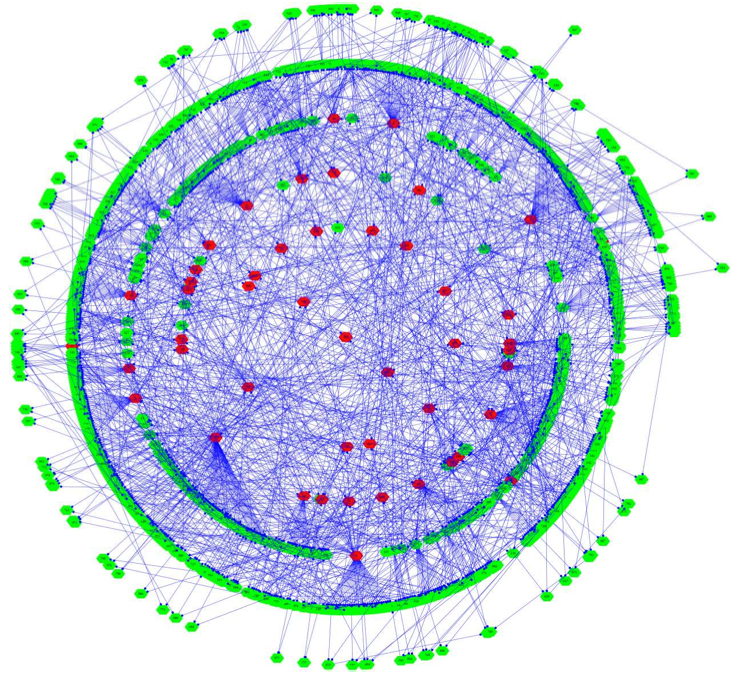}
\caption{A snapshot of the status of the scale-free network at a given time during the infection: infected (red) and susceptible (green) nodes.  The scale-free 
network is made of $N=10^3$ nodes, represented via the $k$-shell decomposition, the $6$-th shell (innermost) and the first one (outermost)~\cite{kitsakNP10,software}.}
\label{picture}
\end{figure}
We apply the fractal analysis to different models of information spreading 
in scale-free networks and we show that it is possible to discriminate between them. 
In particular we show that the standard SIS model~\cite{BarabasiRMP02}
is an oversimplified model for that purpose, as it does not reproduce the real world scenario and thus has to be rejected. 
The SIS model is very well suited for describing the spreading of viruses and illnesses,
however --as we will show in the following-- the SIS model does not reproduce the rich
behaviour observed in world wide web data. Indeed,  
as we focus on information spreading in social dynamics, the piece of information is a rumour, 
a useful knowledge, a marketing announcement or a political or philosophical idea;  
the infected nodes are spreaders that try to convince the neighbours 
to adopt or follow their suggestions. It is then natural to include in the model the fact that 
more spreaders will most likely be more effective in spreading the rumour than a single one (synergy) 
and that spreaders might have different rate of success (influential nodes): 
an opinion maker idea is more likely to be followed by the whole community, 
market leaders new campaign will spread easily and more successfully than that of 
a unknown brand (normalised with respect to the investment), 
and news launched by an important media will most likely be reported by other media. 
We introduce a modified SIS model, where synergy components in information spreading~\cite{rechePRL11}
and the presence of influential nodes~\cite{kitsakNP10} are naturally 
taken into account: 
We show that the fractal properties of the data series depend slightly from the fine 
details of the network, while they strongly depend on the recovery time of the network nodes.
Finally, we show that the modified SIS model is compatible with real data and thus it might allows to 
infer important properties on the information spreading process, i.e. the recovery time of the nodes.

\section{Model}
Throughout this paper we will consider two different models of information spreading. 
The standard SIS model~\cite{BarabasiRMP02}, where the nodes have two possible states, 
infected and susceptible of infection. At every time step, every node that has an infected
neighbour has a probability $\nu$ to become infected. After a typical time $\mu = 1/\delta$ 
(number of steps) where $\delta$ is the recovering probability, 
the node recovers and become susceptible again. 
In the modified SIS model we introduce here, we identify the influential spreaders with those highly connected,  
and we set the probability to infect proportional to the number of links a node has:
at every step, each infected node will try to infect every neighbour 
with probability $\nu = z/z_{max}$, where $z$ is its coordination number 
(number of links to other nodes), and $z_{max}$ is the highest coordination 
number encountered in the network. 
Moreover, every spreader attempts to infect all his neighbours every time step 
for a typical total time $\mu = 1/\delta$, after that it recovers and becomes susceptible again; 
that is every node will experience a synergy effect towards begin infected proportional to the 
number of infected nodes between its neighbours. 
Notice that influential nodes are influential ``locally" as witnessed by the high coordination 
number, and this definition has in principle nothing to do with that introduced in~\cite{kitsakNP10} 
and based on the $k-$shell decomposition that we will refer at as ``global'' influential nodes. 
Finally, notice that synergy and influential nodes effects are included as a (local) and linear 
function of the number the neighbour infected nodes and coordination number of the spreaders;
and that while the network is not directed (infection can spread in both ways), 
the probability of infection is asymmetric reflecting a natural scenario in 
social dynamics (a symmetric model has recently been studied in~\cite{cheng13}).

Independently from the spreading model that drives the dynamics, we consider 
scale-free networks of sizes $N$ built via preferred attachment with a power-law distribution 
of coordination numbers $P(z) \sim z^{-2.33}$~\cite{BarabasiRMP02}. The nodes are ordered by 
means of the $k-$shell decomposition as follows~\cite{kshell,kitsakNP10}: the nodes 
that can be disconnected from the network cutting a single link belong 
to the first shell (i.e. only for this shell they correspond to those with a single link). 
After eliminating the nodes belonging to the first shell from the network, the nodes that 
have a single link remaining belong to the second shell. 
The procedure is repeated until all nodes are eliminated and the last shell is defined. 
Notice that from the second shell on, the coordination number $z$ of a node and its shell usually do not coincide. 
The infection is injected in a random node (either in the whole network or in a predetermined shell) 
and we monitor the number of infected nodes $N_I$ as a function of (discrete) 
time $t_i= i \cdot \Delta t $, where $\Delta t$ is a typical time scale of the system.
Notice that in this model we have only one free parameter, that is the recovery time 
$\mu$ of each infected nodes.
A typical result of this dynamics is shown in Fig.~\ref{picture}, where the status of 
the network is depicted at a given time and infected and susceptible nodes are marked. 

\begin{figure}[t]
\includegraphics[width=0.45\textwidth,trim = 50mm 25mm 30mm 22mm]{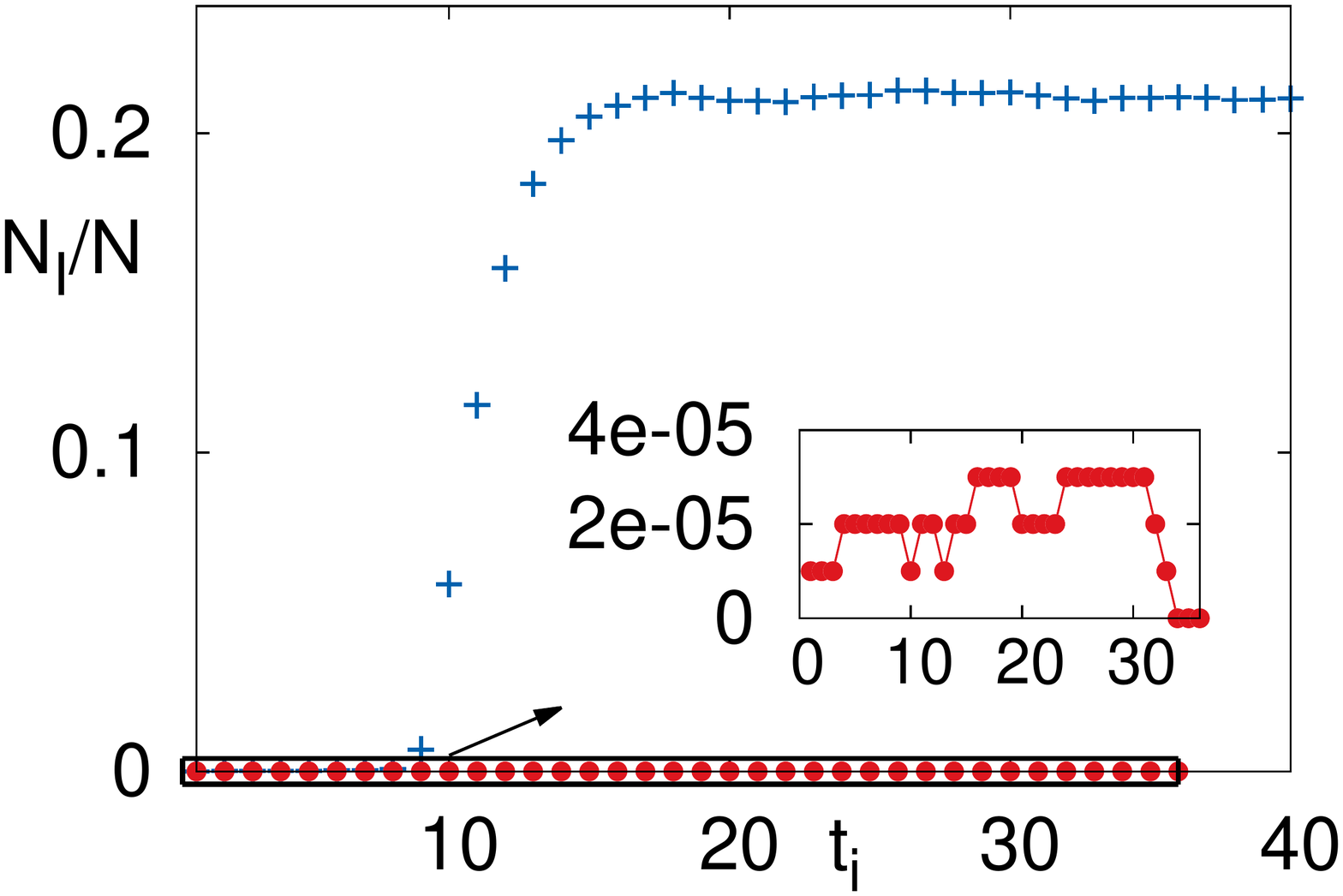}
\includegraphics[width=0.45\textwidth,trim = 13mm 65mm 10mm 70mm, clip]{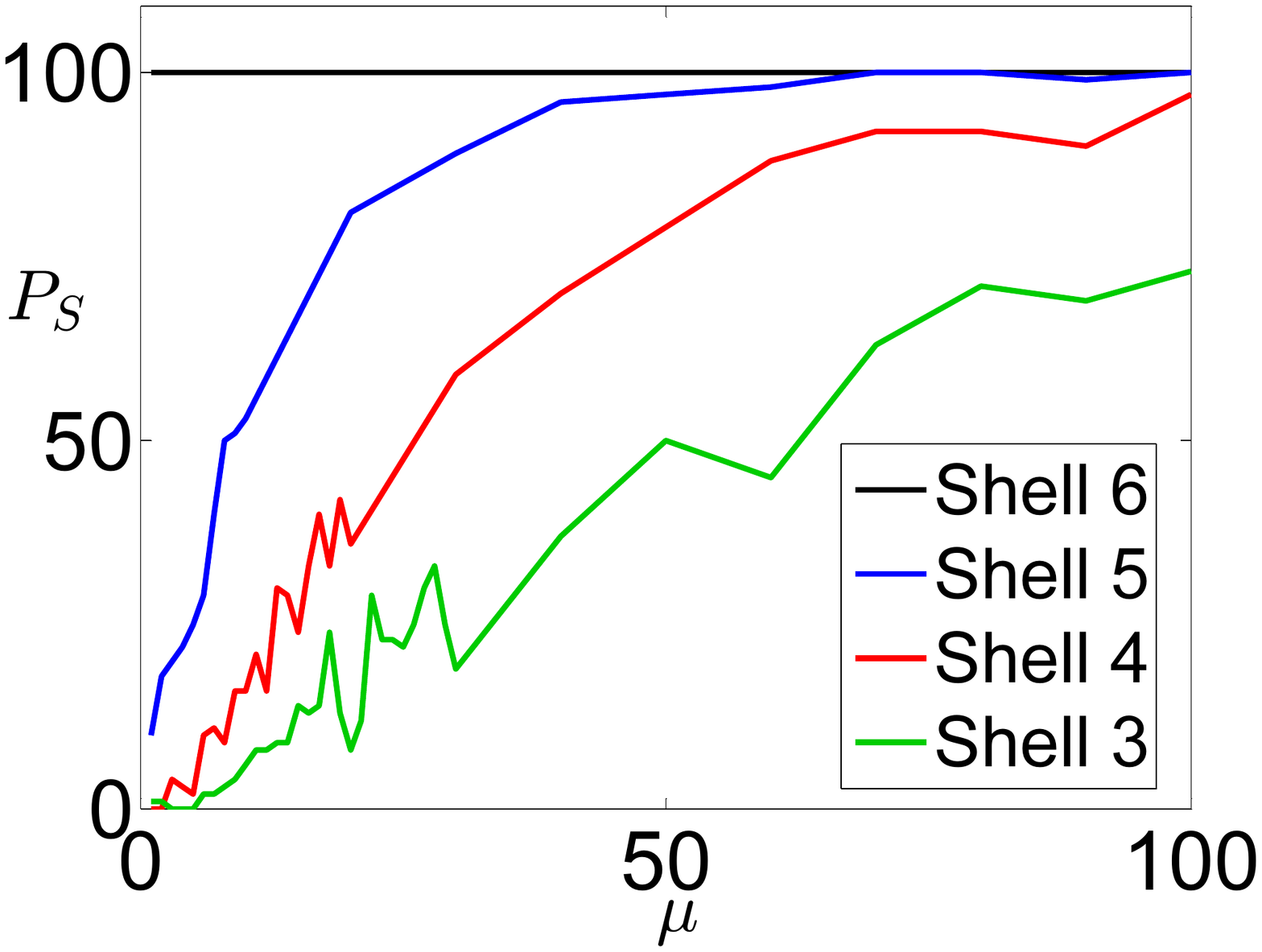}
\caption{Top: Infection level $I(t_i)$ as a function of time for a
successful (blue) and unsuccessful (red) spreading process in the modified SIS model. 
Bottom: Probability $P_S$ for long term persistence of an infection as a function of memorytime $\mu$ 
and shell number.}
\label{timeev}
\end{figure}

\section{Results}
We first focus on the evolution of the infection under the modified SIS model. 
Typical results are shown in Fig.~\ref{timeev} (top), where the percentage of infected network 
$I=N_I/N$ as a function of time is reported for the two different scenarios we have found: It either
dies out very quickly or it becomes permanent and stabilises around a non-zero level.  
{The latter behavior resemble what is found also in other models, see e.g.~\cite{BarthelemyPRL04}}.
This reproduce the common knowledge that ``rumours are hard to die" or the long-standing duration 
of chain letters once they have started. In the bottom panel of Fig.~\ref{timeev} 
we report the probability $P_S$ of having a non dying process as a function 
of the recovery time $\mu$, with starting point in different $k-$shells. 
As can be clearly seen, infections starting in the inner shells have higher probability 
to persist in the network than those started in the outer shells, and 
in general longer the recovery time higher the probability of a permanent infection. 
Notice also that although in the network there is always a node 
with $100 \%$ success spreading rate, i.e. $\nu =1$, this does not implies 
that the infection becomes permanent. That is, a single very connected (and locally influential) node 
cannot deterministically influence the whole network.
Assuming the infection is not started from a poor spreader, 
we however see that the critical recover probability $\delta_c$ above which the infections 
will not be sustainable, scales as $1-\delta_c \propto 1/N$ with the system size (data not shown). 
This behavior is typical for small-world networks~\cite{SatorrasPRL01}.

\begin{figure}[t]
\includegraphics[width=0.48\textwidth, trim = 1mm 125mm 6mm 12mm, clip]{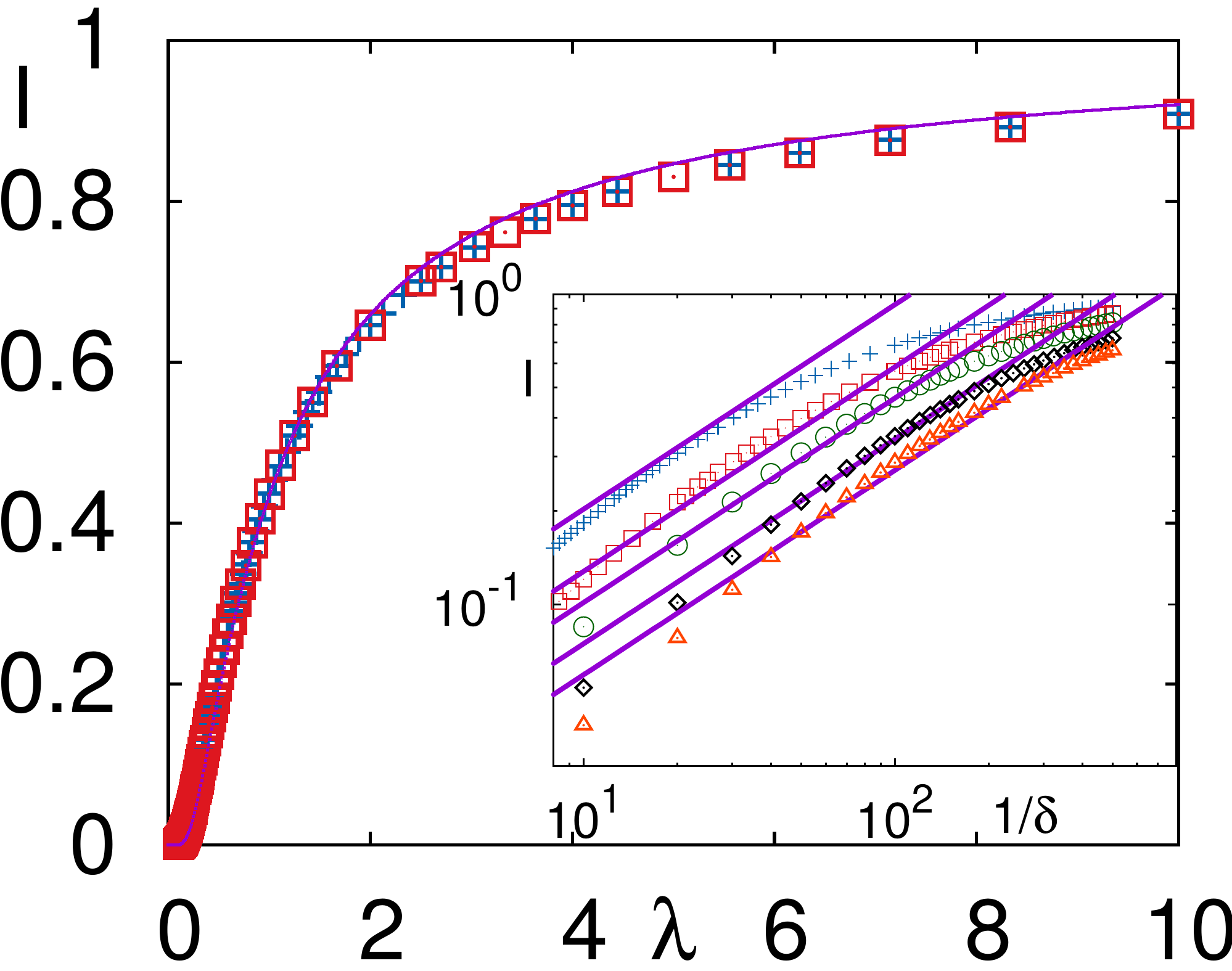}
\caption{Infected ratio of the SIS model (main panel) as a function of $\lambda = \frac{\nu}{\delta}$ and of the modified SIS model as a function of the 
inverse recover ratio $1/\delta$ (inset), 
for different network sizes $N= 10^5, 4 \cdot 10^5, 8 \cdot 10^5, 2 \cdot 10^6, 4 \cdot 10^6$ (blue crosses, red squares, green circles, black diamonds, orange 
triangles). 
Full lines show theoretical lines: Typical s-curve for SIS model and power law for modified SIS model.}
\label{inflev}
\end{figure}

From now on, we concentrate on the cases where the permanent infection occurs, and 
in particular on the average level of infection $\bar I$ and on the time fluctuations 
$\Delta I_i = I_i - \bar I$, where $\bar \cdot$ stands for time average. 
In Fig.~\ref{inflev} we report the infection level $\bar I$ as a function of the 
inverse recovery time $1/\delta$ obtained using the model introduced here and the  
SIS model (as a function of $\lambda = \nu/ \delta$). While the SIS displays the 
typical s-shape dependence~\cite{moreno03}, the modified SIS model results in a 
strictly concave dependence of $\bar I$ with the recovery time $\mu$ until saturation 
effects comes into play.  
As can be seen in Fig.~\ref{inflev}, this dependence is compatible with a power-law scaling (full lines) 
that can be estimated with the following simple theoretical arguments: to sustain 
a permanent infection level, each node has on average to infect another 
node during the recovering time $\mu$, that is the probability of one infection has to be $p_1 \geq 0.5 $.
The probability of infecting one of the $z_i$ neighbours at every step is $\nu_j = z_j/z_{max}$, and from 
elementary probability theory we obtain that the overall probability 
to have at least on successful infection in $\mu$ steps is
\begin{equation}
p_1 = z_j \left[1 - (1- \frac{z_j}{z_{max}})^\nu\right].
\end{equation}
Given that for the majority of nodes $z_i << z_{max}$, we can expand $p_1$ in Taylor series and the 
condition of having a permanent infection ($p_1 \geq 0.5 $) is satisfied by the nodes that fulfil 
\begin{equation}
z^* \geq \sqrt{\frac{z_{max}}{2 \mu}}. 
\label{cond}
\end{equation}
Thus, the number of nodes that can sustain the permanent infection are those 
whose coordination number fulfil Eq.~\eqref{cond} (notice that this condition is 
completely independent from the $k$-shell analysis). 
Their number, that is the infection level $\bar I$ 
can be estimated from the static network properties, as the number of nodes with $z > z^*$ is given by:
\begin{equation}
\bar I = \int^\infty_{z^*} P(z) dz = \frac{C}{(z^*)^{\gamma-1}};
\label{ilev}
\end{equation}
where $C$ is a constant, and $P(z) = z^{ -\gamma}$ 
is the distribution of links of nodes in the network; and in our simulations $\gamma =2.33$. 
From Eq.~\eqref{cond} and Eq.~\eqref{ilev} we obtain
\begin{equation}
\bar I \sim  C \left(\frac{z_{max}}{2}\right)^{-\eta} \cdot \mu^{\eta}, 
\label{estimate}
\end{equation}
where 
\begin{equation}
\eta = (\gamma -1)/2
\label{exponent}
\end{equation}
By means of numerical analysis on different networks and network sizes, we fit the values 
of the constants $z_{max} \sim 2.8 \sqrt{N}$ while $C \sim 2.6$ (data not shown).
The infection rate converges to our theoretical prediction for $\frac{1}{\delta} \geq 10$ as reported in Fig.~\ref{inflev} (inset).
The exponent in the power law from  Eq.~\eqref{estimate} is not obtained by a fit but is deduced from a very basic property of scale free networks; their power law parameter $\gamma$ (see introduction).
Notice that Eq.~\eqref{estimate} is invertible, thus given a network size and an infection level it allows to 
extract the recovery time of the model.

\begin{figure}[t]
\begin{center}
\includegraphics[scale=0.42,trim = 2mm 63mm 10mm 90mm, clip]{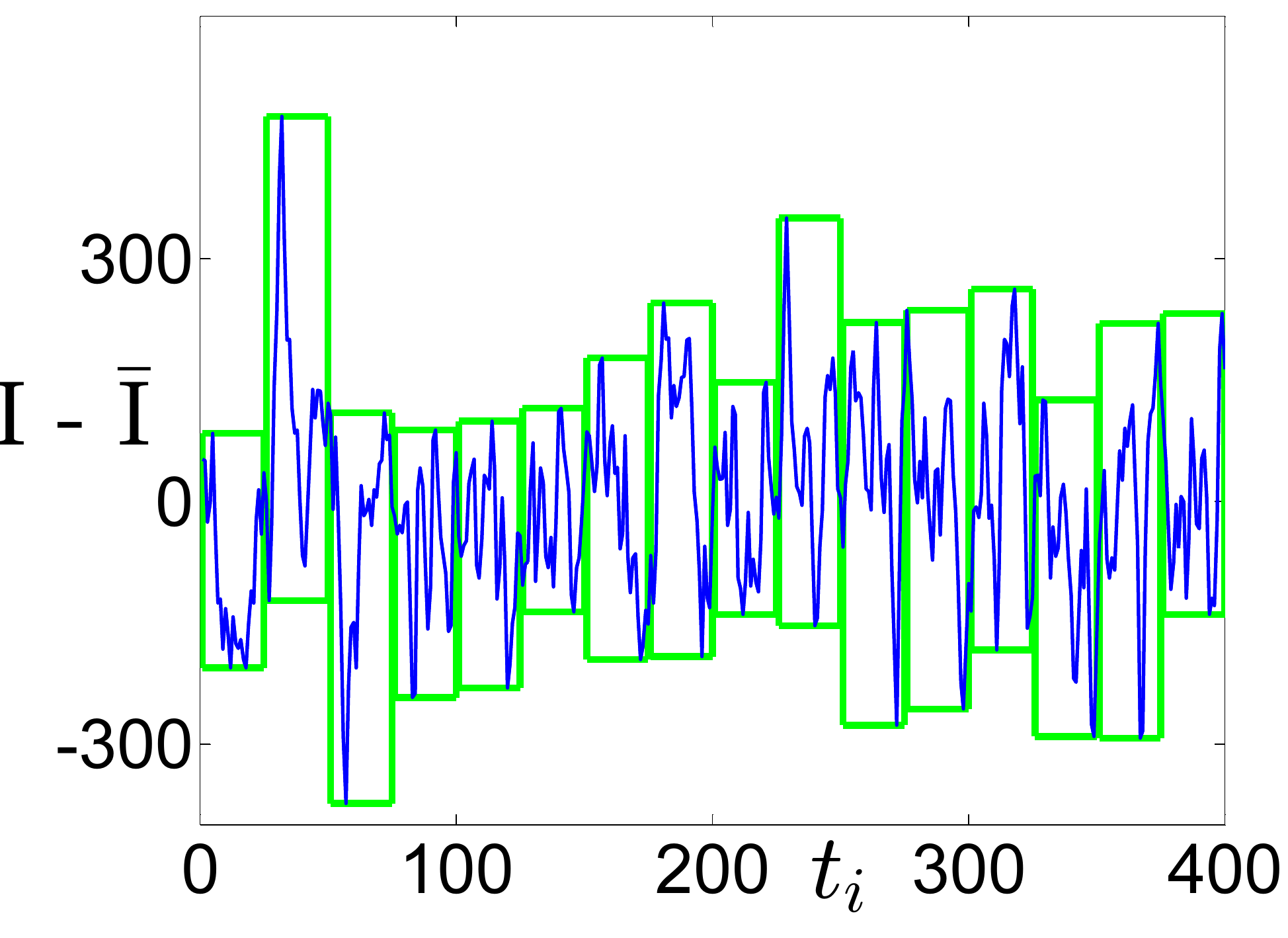}
\caption{A typical time evolution of number of suspensibles I. Fluctuations around their average $\bar I$ are being analysed by grouping them into rectangles 
of equal width $L = t_i- t_j$.}
\label{covering}
\end{center}
\end{figure}
\begin{figure}[t]
\includegraphics[width=0.23\textwidth, trim = 30mm 18mm 40mm 23mm, clip]{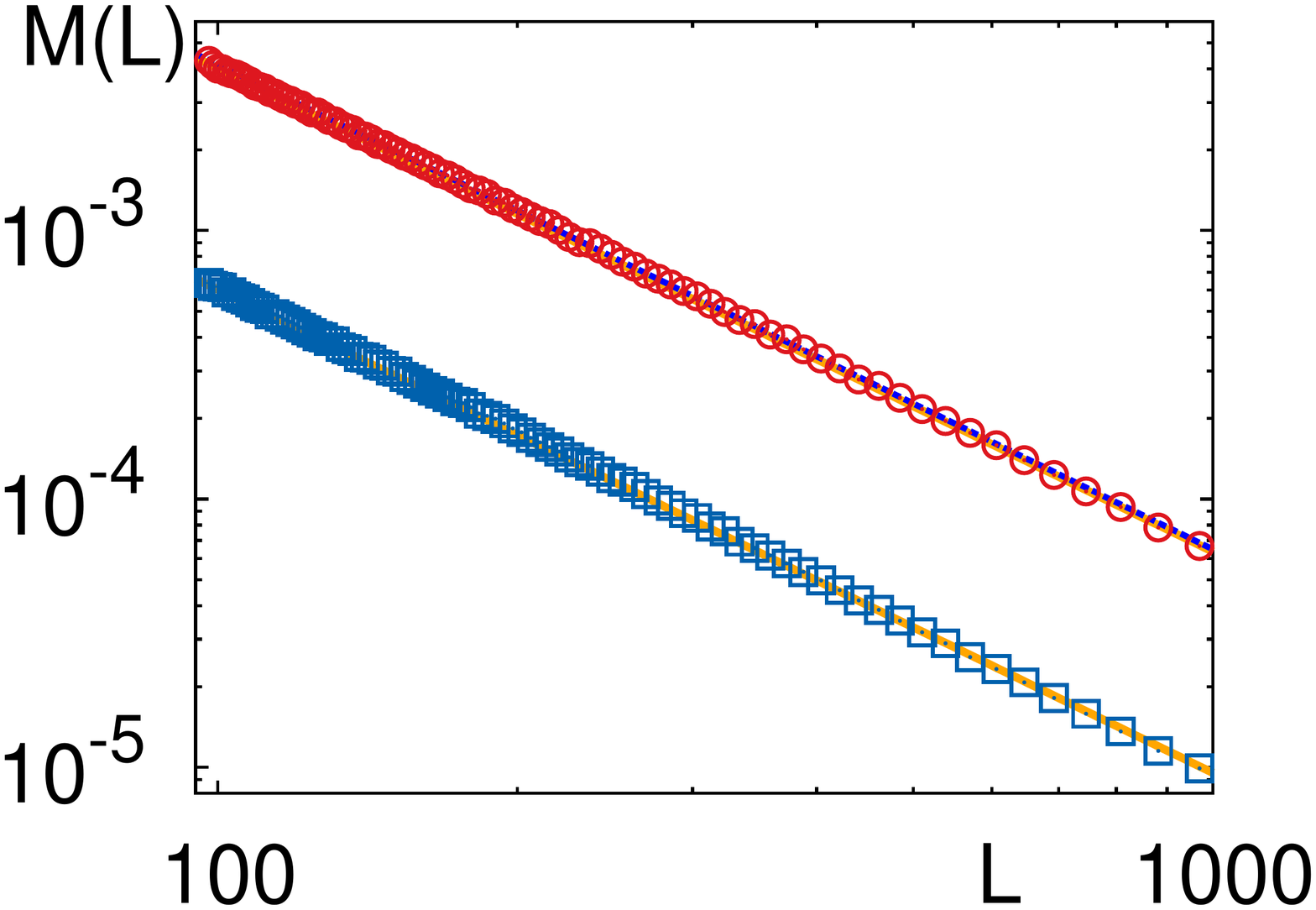}
\includegraphics[width=0.24\textwidth, trim = 30mm 18mm 40mm 23mm, clip]{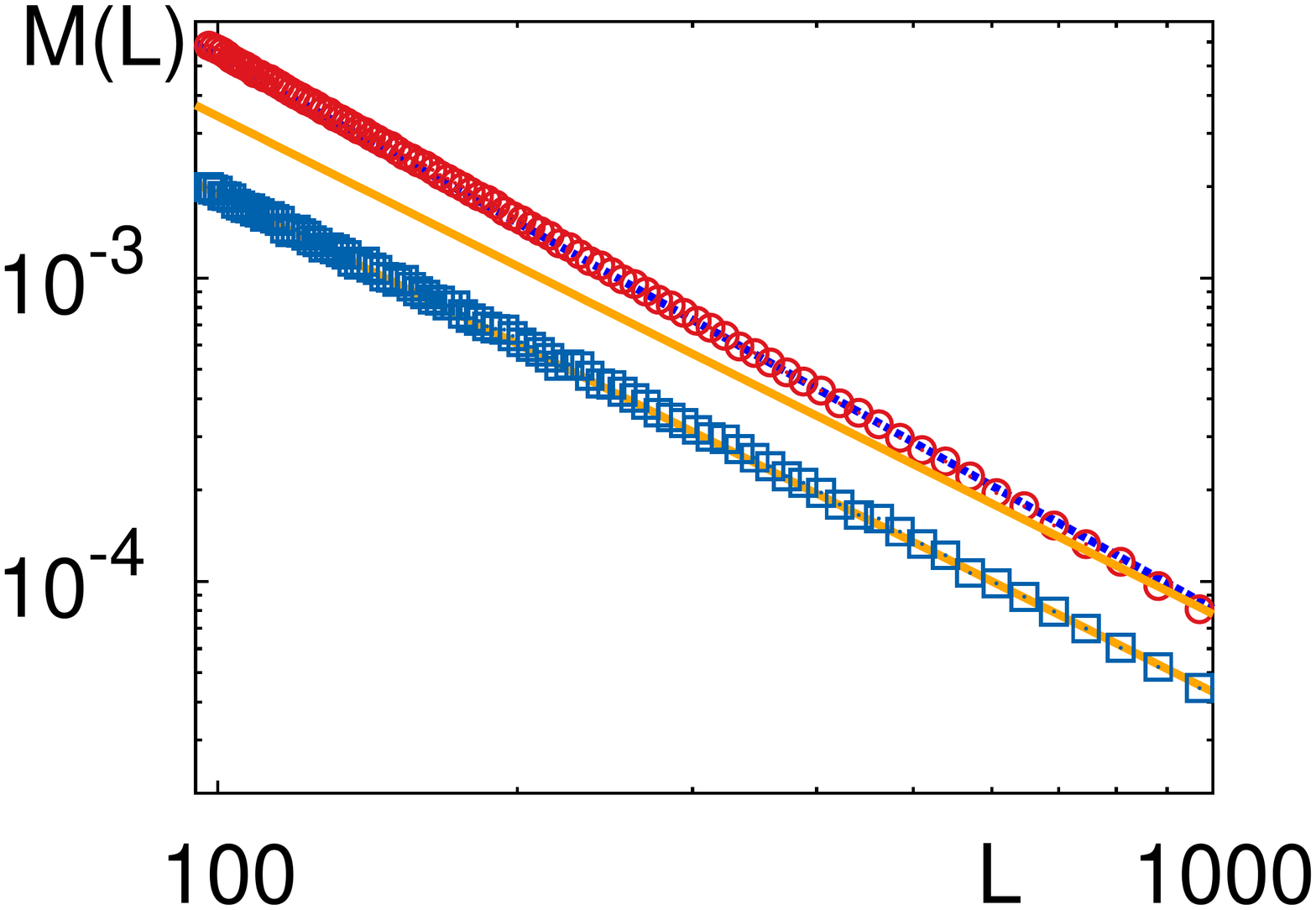}
\includegraphics[width=0.23\textwidth, trim = 45mm 18mm 45mm 23mm, clip]{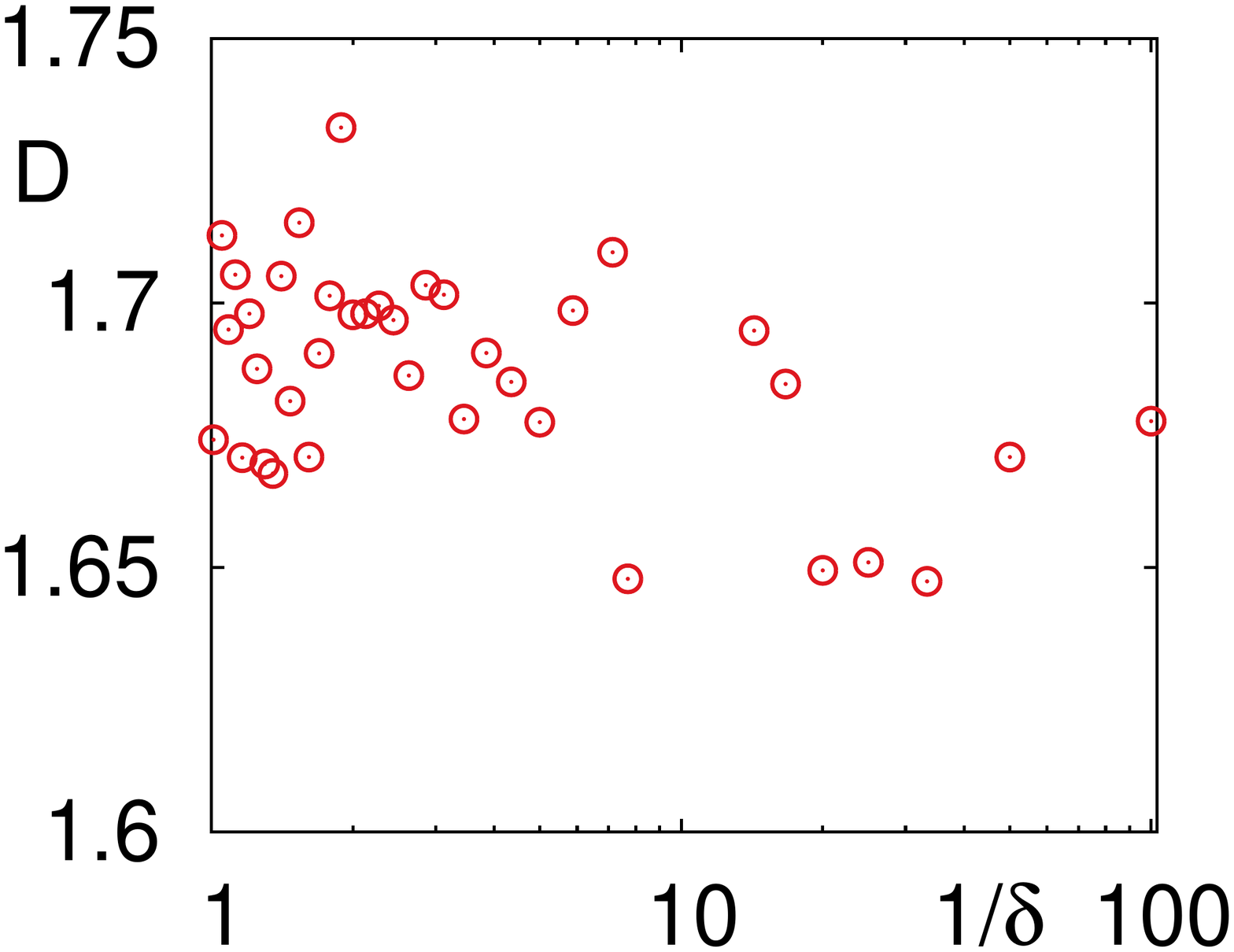}
\includegraphics[width=0.245\textwidth, trim = 40mm 18mm 45mm 23mm, clip]{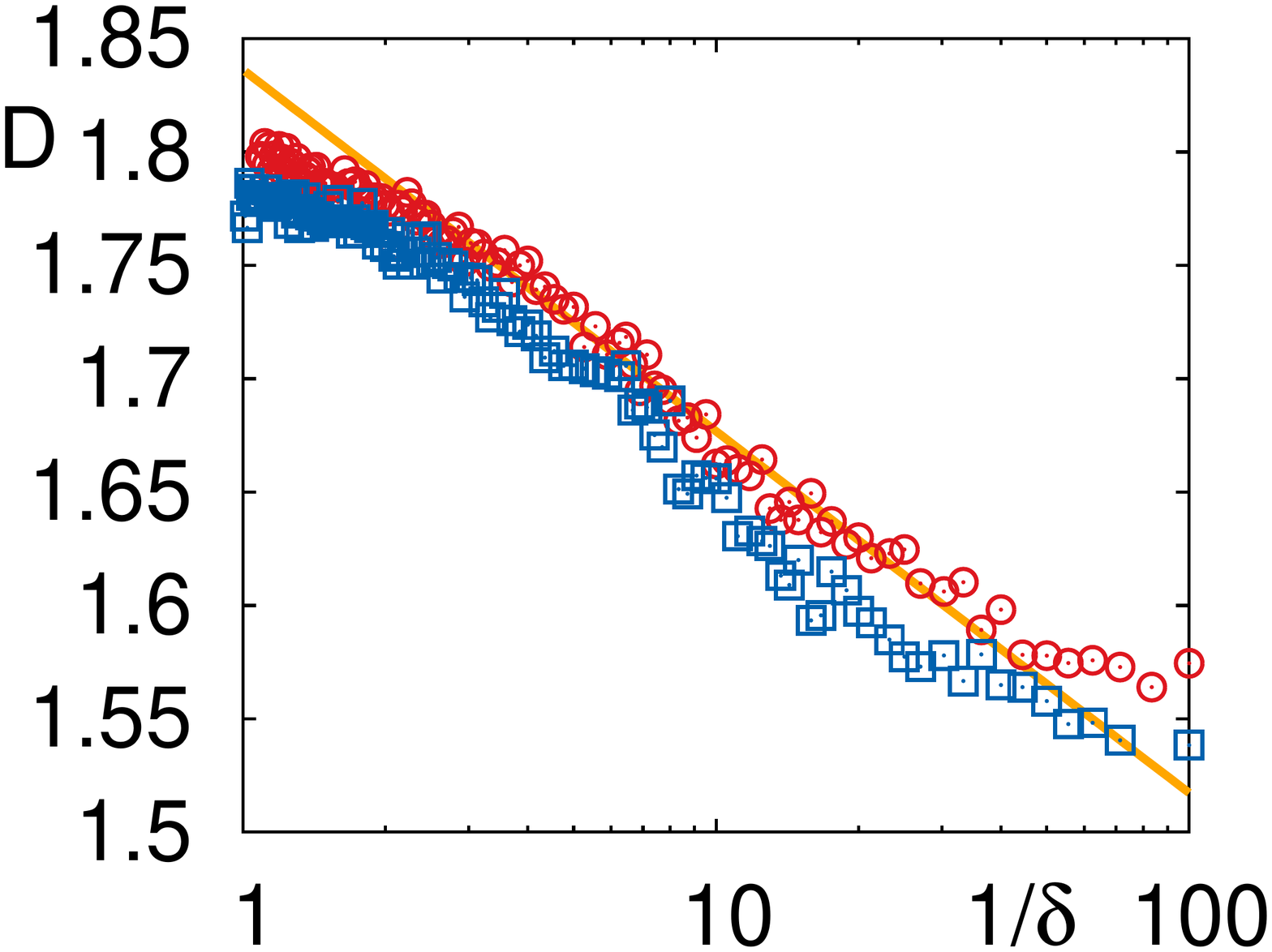}
\caption{
{Top panels show the fitting procedure performed to compute the FD, for the SIS (left) and modified SIS (right) models, 
for two different values of $\delta=0.01, 0.99$ (red circles and blue squares respectively). The full lines are the resulting fitted power laws: in the right panel the lower line is reported also translated for an easy comparison of the two different slopes.}
Lower panels: Fractal dimension of the temporal fluctuations $\Delta I$ for the SIS (left) and modified 
SIS (right) model. Different curves represents different network sizes $N= 10^5, 4 \cdot 10^5 $ (blue squares and red circles respectively). The error on the points is of a few percent. The line shows a fit $f(x) = a+ b \log(x)$ in the 
region $x \in [2:40]$ resulting in $a= 1.829 \pm 0.002$ and $b= -0.154 \pm 0.002$.
}
\label{fractal1}
\end{figure}

We finally, concentrate on the fluctuations of the infection level around the average value $\bar I$, in particular we analyse the fluctuations 
by means of the fractal analysis as introduced in~\cite{montangero12}. The fractal dimension of a signal can be extracted by means of the 
modified box counting algorithm~\cite{sachrajdaPRL98}: 
In the standard box counting algorithm the fractal
dimension $D$ of the signal is obtained by covering the data with a
grid of square boxes of size $L^2$. The number $M(L)$ of boxes
needed to cover the curve is recorded as a function of the box
size $L$. The (fractal) dimension $D$ of the curve is then defined as 
\beq
	D = - \lim_{L \to 0} \log_L M(L). 
\label{fractal}
\eeq
The modified algorithm follows the same lines but 
uses rectangular boxes of size $ L \times \Delta_i$ ($\Delta_i$ is the largest 
excursion  of the curve in the region $L$). Then, the number 
$M(L) = \frac{\sum_i  \Delta_i}{L}$
is computed. 
Such procedure is illustrated in Fig.~\ref{covering} where a typical signal $\Delta I_i$ 
is processed. For any curve a region of box lengths $L_{min} 
< L < L_{max}$ exists where $M \propto L^{-D}$. Outside this region one
either finds $D=1$ or $D=2$: The first equality ($D=1$) holds for
$L<L_{min}$ and it is due to the coarse grain artificially introduced
by any discrete time series. The second one ($D=2$) is obtained for
$L>L_{max}$ and it is due to the finite length of the analysed time
series. The boundaries $L_{min},L_{max}$ have to be chosen properly for
any time series, 
and a power-law fit allows to extract the fractal dimension $D$.
The upper panels of Fig.~\ref{fractal1}  show typical results of this procedure.
The fractal dimension measure the degree of correlations in a time series, as shown for example in~\cite{corrD}: 
In the case of a stationary Gaussian random process 
one can show that if the correlations in the time series are such that $C(h)=1-|h|^\beta$as $h \to 0$ for some 
$\beta \in (0,2]$, then the fractal dimension is related to the exponent as $D=2  - \beta/2$.
Thus, the faster the decaying of the correlation the lower the fractal dimension is: for example 
$D = 2$ corresponds to the case of very slow decaying correlations $\beta = 0$, while $D = 1$ 
to fast decaying correlations $\beta = 2$.

We performed the aforementioned analysis for a wide range of different evolutions, 
for the SIS and the modified SIS model, for different recovery times $\mu$. The 
results are presented in Fig.~\ref{fractal1} (lower panels), where we show that the SIS model 
gives fractal dimension almost constant of about $D_{SIS} \sim 1.7 \pm 0.05$. On 
the contrary, the modified SIS model results in a richer behaviour: the fractal 
dimension gives values in the range $[1.52 \pm 0.05 : 1.85 \pm 0.05]$ and 
scales approximatively as 
\begin{equation}
D(1/\delta) - 1.83 = -0.154 \cdot \log_{10}(1/\delta).
\label{Ddelta} 
\end{equation}
This result is slightly influenced by 
the system size and thus allows in principle, to extract important information on real world 
data sets. 

\section{WWW data analysis} 
As already shown in~\cite{montangero12}, the fractal dimension analysis can be 
performed from the time evolution of web pages that include some keywords.
{Having a proper model of the processes behind the spreading process 
it might be possible to obtain from the fractal dimension time evolution some quantitative 
measurement of important system parameters. Indeed, 
the fractal dimension analysis results could give an insight 
on the underlying dynamics generating the overall signal (or the structure and social behaviour 
of the communities under study if one wants to put forward the hypothesis). 
For example, in our case, 
if the process is correctly modelled by our modified SIS model, 
one could calculate the memory time $\mu$ of the spreaders, from Eq.~\eqref{Ddelta}.
A typical example of such analysis is presented in in Fig.~\ref{www} where 
the time serie, sampled every hour, is reproduced and the correspondent 
fractal dimension computed spans from $D=1.2 \pm 0.2$ to $D=1.7 \pm 0.2$. 
A first clear results of this analysis is that the SIS model has to 
be rejected as it cannot reproduce the real data properties, indeed the 
real process is characterised also by fractal dimensions far away from $D_{SIS}$. 
On the other side, the modified SIS model allows instead to better reproduce most of 
the experimental data.
Indeed, working under the assumption that the main features are grasped 
by the modified SIS model, from the fractal analysis 
of the signal one can infer an average 
recovery time of the process. As a result, one might distinguish processes that are characterized 
by different time-scales that spams from a few hours to a few days.
Note that during the Christmas holidays the fractal dimension is increased, which signals a very frequent 
activity and correspondingly, from Eq.~\eqref{Ddelta}, the shortest recovery time and hence reflecting an increased importance. After that period the fractal 
dimension lowers again which means a return to usual interest.
\begin{figure}[t]
\includegraphics[width=0.5\textwidth, trim = 43mm 20mm 35mm 50mm, clip]{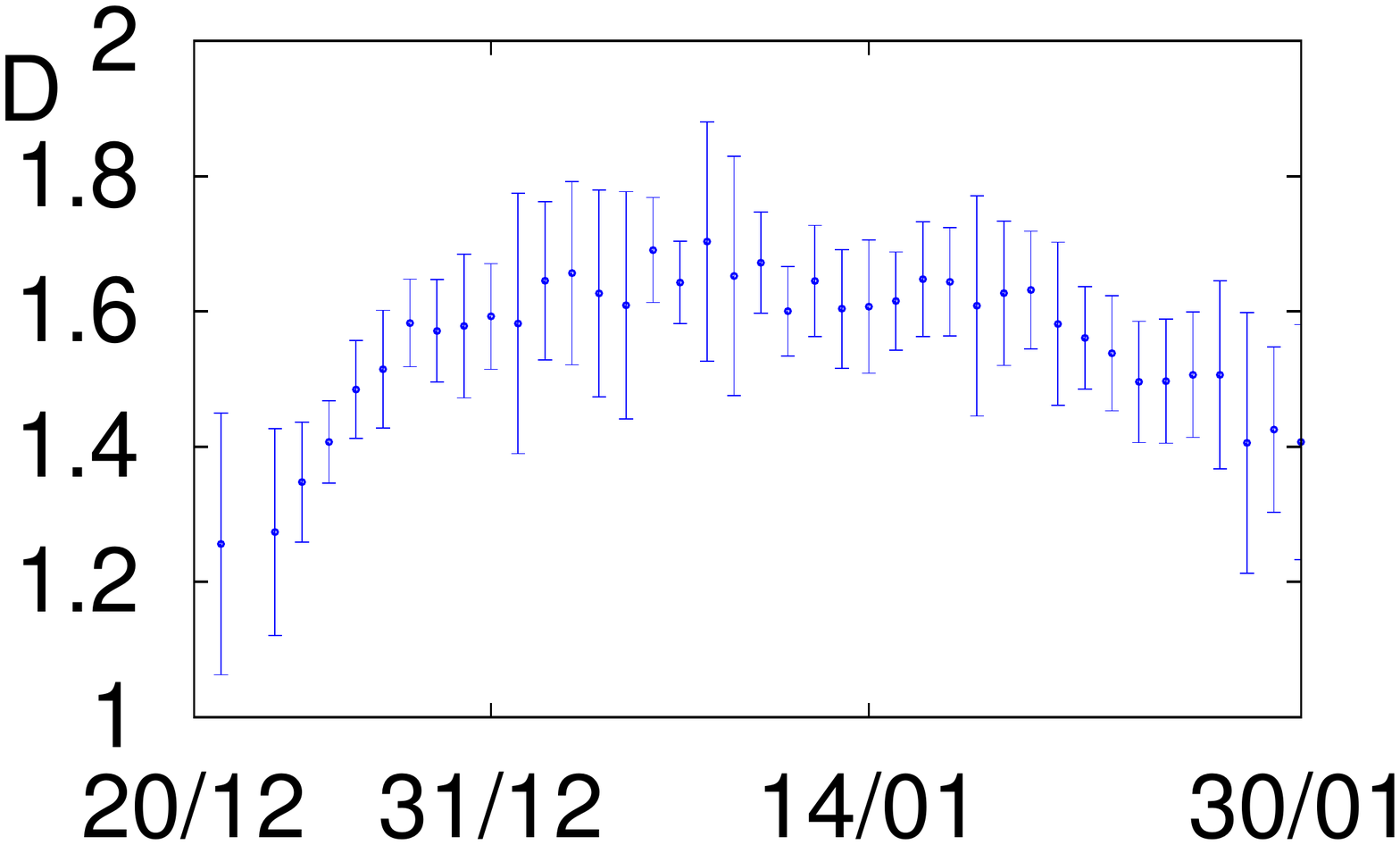}
\caption{Fractal dimension of the temporal fluctuations $\Delta I$ for Myspace computed with the data from December 2008 
until January 2009 and a sampling frequency of one hour. }
\label{www}
\end{figure}

\section{Conclusions}
We have shown that tools from signal processing analysis and in particular the fractal analysis might be used 
to test the theoretical models used to describe networks dynamics. 
In particular, we have compared the possible results from a SIS model 
and shown that it does not allow for a wide enough variation of fractal dimension, thus failing to describe experimental data. 
We have then introduced a modified SIS model to better match the real-world data, obtaining an improved description even though not completely satisfactory. 
{There is still some discrepancy between the real-world data analysis and the results from the 
modified SIS model introduced here. This might be due to the fact that the infection 
probability does not exactly scale linear with the number of neighbors as we have assumed. 
Indeed, a single parameter model might not be sufficiently versatile to describe quantitatively complex social dynamics.  
For example, including contrary or multiple opinions as shown in~\cite{Schwaemmle07} might also improve the descriptive power of the model.} Finally, the 
discrepancy between the real-world data analysis and the results from the modified SIS model introduced here call for further 
developments: extended search for a perfect matching model and a comparison with a statistically significant real-world data will be subject of future work. 

 \begin{acknowledgments}
We thankful acknowledge M.Furini for support in the data harvesting and 
we thank the bwGRiD for the computational resources. 
\end{acknowledgments}


\begin{thebibliography}{10}

\bibitem{BarabasiNP11} A.-L. Barab\'asi, Nature Phys. {\bf 8}, 14 (2011).

\bibitem{BarabasiRMP02} A.-L. Barab\'asi,  and R. Albert, Rev. Mod. Phys. {\bf 74}, 47 (2002).

\bibitem{castellano09}  C. Castellano, S. Fortunato, and V. Loreto, Rev. Mod. Phys. {\bf  81}, 591 (2009). 

\bibitem{leskovec07} J. Leskovec, L. A. Adamic, and B. A. Huberman, ACM Transactions on the Web {\bf 1}, 1 (2007).

\bibitem{lewisPNAS12} K. Lewis, M. Gonzalez, and J. Kaufman, Proc. Natl. Acad. Sci. USA {\bf 109}, 68 (2012).

\bibitem{Liben-nowell2008} D. Liben-nowell and J. Kleinberg, Proc. Natl. Acad. Sci. USA {\bf 105}, (2008).

\bibitem{viswanath09} B. Viswanath, A. Mislove, M. Cha, and K. P. Gummadi, 
Proceedings of the 2nd ACM Workshop on Online Social Networks, 37 (2009).

\bibitem{pala12} C. Palazuelos and M. Zorrilla, Proceedings of the 2012 Joint EDBT/ICDT Workshops,  9 (2012).

\bibitem{Lind07} { P. G. Lind, L. R. da Silva, J. S. Andrade and H. J. Herrmann, Phys. Rev. E {\bf 76}, 036117 (2007).}

\bibitem{Gallos12} { L. K. Gallos, P. Barttfeld, S. Havlin, M. Sigman and H. A. Mokse, Sci. Rep. {\bf 2}, 454 (2012).}

\bibitem{centolaSc10} D. Centola, Science {\bf 329}, 1194 (2010).

\bibitem{moreno04} Y. Moreno, M. Nekovee, and A. F. Pacheco, Phys. Rev. E {\bf 69}, 066130 (2004).

\bibitem{youngPNAS11} H. P. Young, Proc. Natl. Acad. Sci. USA {\bf 108}, 21285 (2011).

\bibitem{rechePRL11}  F. J. Porez-Reche, J. J. Ludlam, S. N. Taraskin, and C. A. Gilligan, Phys. Rev. Lett. {\bf 106}, 218701 (2011).

\bibitem{shaoPRL09} J. Shao, S. Havlin, and H. Stanley, Phys. Rev. Lett. {\bf 103}, 018701 (2009).

\bibitem{Liu2012} C. Liu and Z. Zhang, Comm. Nonl. Science and Numerical Simulation, {\bf 19},  896 (2014).

\bibitem{Araujo10} { N. A. M. Araujo, J. S. Andrade, H. J. Herrmann, Plos One 5, e12446 (2010).}

\bibitem{Moreira06}{ A. A. Moreira, D. R. Paula, R. N. Costa Filho and J. S. Andrade, Phys. Rev. E {\bf 73}, 065101 (2006).}

\bibitem{hbp} http://www.humanbrainproject.eu


\bibitem{grigo10} {\it ``Complex Webs: Anticipating the Improbable"},  B. West and P. Grigolini, Cambridge University Press (2010).


\bibitem{montangero12} S. Montangero and M. Furini, 
US Patent, Serial No.: 12/635,004. M. Furini, S. Montangero, OJWT in press. 

\bibitem{kitsakNP10}
M. Kitsak, L. K. Gallos, S. Havlin, F. Liljeros, L. Muchnik, H. E. Stanley, and H. a. Makse, Nature Phys.  {\bf 6}, 888 (2010).


\bibitem{cheng13} J.-J. Cheng, Y. Liu, B. Shen, and W.-G. Yuan,  European Physical Journal B {\bf 86}, 29 (2013).


\bibitem{kshell} 
B. Bollobos, Proceedings of the Cambridge Combinatorial Conference in Honor of P. Erd\"os Vol. 35 (Academic, 1984);
S. Carmi, S. Havlin, S. Kirkpatrick, Y. Shavitt, and E. Shir, 
Proc. Natl. Acad. Sci. USA {\bf 104}, 11150  (2007).

\bibitem{BarthelemyPRL04} { M. Barthelemy, A. Barrat, R. Pastor-Satorras and A. Vespignani, Phys. Rev. Lett.  {\bf 92}, Number 17 
(2004).}

\bibitem{SatorrasPRL01} { R. Pastor-Satorras and A. Vespigani, Phys. Rev. Lett.  Vol. 86, Number 14 (2001).}

\bibitem{moreno03} Y. Moreno and A. Vazquez, Eur. Phys. J. B {\bf 31}, 265 (2003).

\bibitem{sachrajdaPRL98}  A. S. Sachrajda, R. Ketzmerick, C. Gould, Y. Feng, P. J. Kelly, A. Delage, and Z. Wasilewski
Phys. Rev. Lett.  {\bf 80}, 1948, (1998).

\bibitem{corrD} T. Gneiting and M. Schlather, SIAM Review, {\bf 46},  269 (2004). 
 
\bibitem{software}  We acknowledge the graphical open source software Graphviz (http://www.graphviz.org).  

\bibitem{Schwaemmle07} {V. Schwaemmle, M.C. Gonzalez, A. A. Moreira, J.S. Andrade and H. J. Herrmann, Phys. Rev. E {\bf 75}, 066108 (2007). }




\end{thebibliography}

\end{document}